\documentclass[twocolumn,pra,tighten,floatfix,showpacs]{revtex4-1}
\usepackage{graphicx}

\def\<{\langle}
\def\>{\rangle}

\def\be{\begin{equation}}
\def\ee{\end{equation}}

\begin{document}
\preprint{cond-mat} \title{Electoral Susceptibility}

\author{G. C. Levine, B. Caravan and J. E. Cerise*}

\address{Department of Physics and Astronomy, Hofstra University,
Hempstead, NY 11549 \\
*Department of Biostatistics, Mailman School of Public Health, Columbia University, NY 10032}

\date{\today}

\begin{abstract} In the United States electoral system, a candidate is elected indirectly by winning a majority of electoral votes cast by individual states, the election usually being decided by the votes cast by a small number of "swing states" where the two candidates historically have roughly equal probabilities of winning.  The effective value of a swing state in deciding the election is determined not only by the number of its electoral votes but by the frequency of its appearance in the set of winning partitions of the electoral college. Since the electoral vote values of swing states are not identical, the presence or absence of a state in a winning partition is generally correlated with the frequency of appearance of other states and, hence, their effective values.  We quantify the effective value of states by an {\sl electoral susceptibility}, $\chi_j$, the variation of the winning probability with the "cost" of changing the probability of winning state $j$.  We study $\chi_j$ for realistic data accumulated for the 2012 U.S. presidential election and for a simple model with a Zipf's law type distribution of electoral votes.  In the latter model we show that the susceptibility for small states is largest in "one-sided" electoral contests and smallest in close contests.  We draw an analogy to models of entropically driven interactions in poly-disperse colloidal solutions.

\end{abstract}

\maketitle
\section{Introduction} 

The President of the United States is indirectly elected by the electoral college; the electoral college consists of 538 electors, with each state apportioned an amount equal to the amount of Senators and Representatives from that state serving in Congress. In order to win the Presidential election, the candidate must receive at least 270 electoral votes.

With the exception of two states (Nebraska and Maine which apportion their votes by congressional district), each state awards the whole sum of their electoral votes to the candidate who wins the popular vote in that state. Because of the discrete nature in which votes are awarded, there are $2^{51}$ ways in which the election could go.
 
However, using polling data and recent election history, we can determine with great certainty that many states will always go Democrat or Republican. Thus, the outcome of the election is determined by a small number of "swing states" which poll nearly evenly for both parties. Using recent polling data, a typical list might be: Colorado (9), Florida (29), Iowa (6), Michigan (16), Nevada (6), New Hampshire (4), North Carolina (16), Ohio (18), Virginia (13), and Wisconsin (10). (Each state is listed with its electoral votes in parenthesis.) In this instance, the Democratic candidate enters the swing state contest with 221 electoral votes and the Republican with 191 votes; either candidate needs 270 votes to win the election. Only counting these ten states, there are $2^{10} =1024$ partitions of the electoral college. 

Of the 1024 possible partitions of the swing state electoral college, 741 of them achieve 49 or more electoral votes thus leading to a Democratic victory, and 283 of them achieve 79 or more electoral votes leading to a Republican victory. The value of a state to a candidate has not only to do with its electoral size, but its "fit" with other states as measured by the frequency it appears in winning partitions. To illustrate this interaction between states, if the Democrat loses Ohio and Florida, he can still win the election with Michigan, Wisconsin, Virginia, New Hampshire and Nevada or Iowa, achieving exactly the $49$ necessary votes.  However, losing Virginia makes New Hampshire, Nevada and Iowa much less valuable.

Now consider the likelihood that a specific partition would occur, assuming $p_a, p_b \ldots$ are the probabilities that the Democrat will win states $a,b, \dots$, and $1-p_c, 1-p_d, \ldots$ are the probabilities of losing state $c,d,\ldots$.  The probability that the election follows a particular partition (call it $g$) is then,

\begin{equation}
P_{g}=p_a p_b \ldots (1-p_c) (1-p_d) \ldots
\end{equation}
Thus, we have a set of probabilities, $\{P_g\}$, corresponding to the 1024 different scenarios for the outcome of the swing states. Using this, the probability of the Democrat winning, $E$, is:

\begin{equation}
E = \sum_{{\rm partitions} \,\,g} P_g \ \theta(N_g - N_W)
\end{equation}
where $N_g$ is the electoral votes awarded to the candidate in partition $g$, and $N_W$ is the number of electoral votes the candidate needs from the swing states to win. The unit step function, $\theta$, restricts the sum to winning partitions.
 
Given the discrete nature of the electoral college and inhomogeneous electoral vote values, we pose the following question: if $s$ votes for one candidate within a particular state could be added for a set "cost"---thereby increasing the probability of winning that state---in which state should that investment be made? \cite{ev,silver,mebane}  We define this quantity by an {\sl electoral susceptibility} $\delta E/\delta s_j$ which roughly measures the variation of the winning probability with the probability of winning state $j$. 

In an inhomogeneous electoral college, the removal of a single state $i$ affects the number of times state $j$ appears in the set of winning partitions. This is true in general for any pair of states $(i,j)$ and leads to unintuitive correlations between the states. Associating entropy with the logarithm of the number of appearances of a state within the set of winning partitions, the entropy per state (in effect, the chemical potential) is not additive and the states may be said to "interact." Analogously to thermodynamics, the susceptibility to a change in an external parameter involves the variation of free energy with respect to the parameter. Within this analogy, the free energy cannot be written as a sum of individual particle free energies because of interactions between the particles. 

This state of affairs is closely related to the phenomenon of entropically driven attraction in bi-disperse colloidal solutions. In solution, the entropy of the small particles depends implicitly on the coordinates of the large particles. The entropy is then maximized by configurations where the large particles are close to one another reducing the volume they exclude.

\section{electoral probability and susceptibility}

In our model let there be $M$ swing states having electoral vote values $\{ N_j\}$ and winning probabilities for candidate $A$ of $\{ p_j\}$, for $j=1\ldots M $.  Consider a partition to be defined as a set of integers $\{i_1\ldots i_M \}$, with $i_j=0,1$.  The probability of a partition is
\begin{equation}
\label{part_prob}
p(i_1\ldots i_M) = \prod_{j=1}^M p_j^{i_j} (1-p_j)^{1-i_j}
\end{equation}
and the probability of candidate $A$ winning an election that requires $N_w$ electoral votes from the swing states is then:
\begin{equation}
\label{prob}
E(N_w) = \sum_{ i_1,\ldots, i_M=0,1} {p(i_1\ldots i_M) \theta(\sum_{j=1}^M{i_j N_j-N_w})}
\end{equation}

Let $s_j$ be the number of additional votes for $A$ acquired in state $j$, expressed in terms of equivalent electoral votes. We are interested in computing the following quantity which we define as the Electoral Susceptibility (ES)
\begin{equation}
\chi_j \equiv \frac{\partial E(N_w)}{\partial s_j}=\frac{\partial E(N_w)}{\partial p_j}\frac{\partial p_j}{\partial s_j}
\end{equation}
The first factor will turn out to measure an unormalized combinatoric increase in $E$ with inclusion of state $j$.  The second factor is dominant when the probabilities $p_j$ deviate strongly from $1/2$ and the state $j$ is no longer strictly a swing state. To express the susceptibility, note that $E(N_w)$ may always be written in terms of conditional probabilities based upon state $j$:
\begin{equation}
\label{conditional}
E(N_w) = p_j E_j(N_w - N_j) + (1-p_j)E_j(N_w)
\end{equation}
where $E_j(K)$ is the probability of winning $K$ electoral votes without state $j$. Therefore 
\begin{equation}
\label{dconditional}
\chi_j = (E_j(N_w - N_j) -E_j(N_w))\frac{\partial p_j}{\partial s_j}
\end{equation}
It is now seen that the first factor encodes the correlations that are present for a particular $N_w$ by measuring how many additional winning partitions are available with the inclusion of state $j$.

\begin{figure}[ht]
\includegraphics[width=7.5cm]{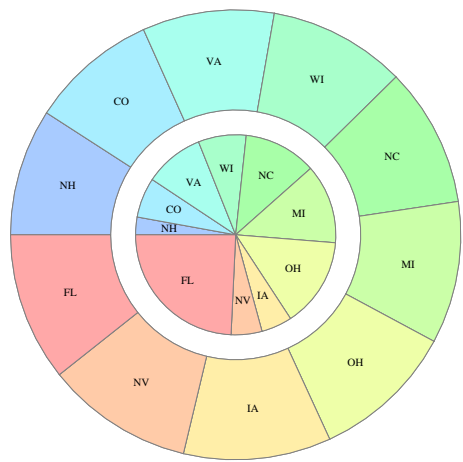}
\caption{\label{pequalshalf} (Color online) Electoral susceptibilities, $\chi_j$, for $N_w=49$ and $p_j =1/2$ displayed in relative terms (outer pie). The small differences between the states reflect interactions, or correlations between their appearance in the set of winning partitions. Electoral susceptibility without normalization, $\partial E/\partial p_j (=N_j \chi_j)$, displayed in relative terms (inner pie.)}
\end{figure}

\begin{table}[h]
\begin{tabular}{|lrr|}
\hline
& $p_j=1/2$ & \\
State          & $\chi_j$ & $\chi_j / \sum$ \\ 
\hline 
Florida	&	0.0154	&	0.107	\\
Nevada	&	0.01530	&	0.106	\\
Iowa	&	0.01530	&	0.106	\\
Ohio	&	0.01487	&	0.103	\\
Michigan	&	0.01477	&	0.102	\\
North Carolina	&	0.01445	&	0.100	\\
Wisconsin	&	0.01426	&	0.099	\\
Virginia	&	0.01367	&	0.095	\\
Colorado	&	0.01324	&	0.092	\\
New Hampshire	&	0.01318	&	0.091	\\
\hline
\end{tabular}
\caption{\label{pequalshalf_table}Electoral susceptibilities, $\chi_j$, for $N_w=49$ and $p_j =1/2$. Despite the large differences in the number of electoral votes for each state, the susceptibilities are all very close in value to each other. However, the residual variations in $\chi_j$  between states reflect interactions or {\sl correlations} between two or more states' appearance in the set of winning partitions }
\end{table}

To complete the evaluation of this expression we must consider how the individual state probabilities, $\{p_j\}$ are formed from polling data. For example, If candidate $A$ is ahead in state $j$ by $x_j = 0.06$ $(6\%)$, a reasonable probability distribution for candidate $A$'s fractional lead, $x$, given polling variance, $\sigma_j^2$ is the distribution:
\begin{equation}
\label{normal}
\tilde{p}_j(x) = \frac{1}{\sqrt{2\pi \sigma_j^2}} \exp(-\frac{1}{2}\frac{(x-x_j)^2}{\sigma_j^2})
\end{equation}
The probability, $p_j$, of candidate $A$ winning state $j$ is then
\begin{equation}
p_j = \langle 1 \rangle_+ \equiv  \int_0^\infty {\tilde{p}_j(x)  dx} 
\end{equation}

The additional acquired electoral votes, $s_j$, effect a shift in $x_j \rightarrow x_j + \delta x_j$, where $\delta x_j = s_j/N_j$. It is possible to show that
\be
\label{dpds}
\frac{\partial p_j}{\partial s_j}|_{s_j = 0} = \frac{1}{N_j \sigma_j^2} \langle x-x_j\rangle_+
\ee
When state $j$ is a pure swing state ($x_j = 0$), $\langle x-x_j\rangle_+ = \sigma_j/\sqrt{2\pi}$; the ES may then be compactly written
\begin{equation}
\chi_j = \frac{1}{\sqrt{2\pi}}(E_j(N_w - N_j) -E_j(N_w))\frac{\alpha_j}{N_j \sigma_j}
\end{equation}
The coefficient $\alpha_j \equiv \sqrt{2\pi}\langle x-x_j\rangle_+/\sigma_j$ has a maximum value of unity and expresses the reduction of the ES as the probability of winning state $j$ deviates from $1/2$. 

When $\alpha_j =1 $ (pure swing state) we see that the susceptibility is "normalized" by a factor $1/N_j$.  This is (constitutionally) reassuring in that we expect the impact of a state on the winning probability, $\partial E/\partial p_j$, to be proportional to $N_j$ so that the election is "fair,"  at least in the limit of an infinite electoral college.   Thus the susceptibility as defined is an electoral per capita susceptibility and should be approximately constant by the central limit theorem.  We demonstrate this feature in the next section when computing the susceptibilities for the electoral college when $p_j =1/2$. 

It is interesting to note that the susceptibility $\chi_j$, for state $j$, is identical for both candidates even if  $N_w$ is very small or large (a "one-sided" contest), and even if states deviate strongly from pure swing state ($p_j = 1/2$) status.   Denoting the candidate $A$ ($B$) susceptibility and probability, $\chi^{A(B)}_j$ and $E^{A(B)}$, respectively, it can be shown that
\be
\chi^B_j = \frac{\partial E^B}{\partial p^B_j} \frac{\partial p_j^B}{\partial s_j} =\frac{\partial E^A}{\partial p^A_j} \frac{\partial p_j^B}{\partial s_j} 
\ee
because $E^B(p^B_j) = 1 - E^A(1-p^B_j)$.  The second factor in the equation above, ${\partial p_j^B}/{\partial s_j}$, defined in equation (\ref{dpds}), may be shown to be equal to its $A$ counterpart by noting that $\langle x-x_j\rangle_{+} = \langle x+x_j\rangle_{+}$. 

Actually, this argument is only strictly correct with an odd number of swing state electoral votes, that is, a swing state contest with no possible tie.  When the total electoral count, $N_T$, is even (as it is in the example of section IV) there is a slight asymmetry in that if $N_w$ is the winning swing state electoral count for the Democrat, $N_T - N_w +1$ is the winning swing state electoral count for the Republican.

One might also consider a higher rank susceptibility that encodes correlations or {\sl synergy} between states: how is the susceptibility of state $i$ affected by a change in probability of state $j$.  Considering only the equal probability case ($p_j=1/2$) and using the same analysis leading to equations (\ref{conditional},\ref{dconditional}), the variation $\delta \chi_i/\delta p_j$ given by:
\begin{eqnarray}
\frac{\partial^2E}{\partial p_i \partial p_j} & = & [E_{ij}(N_w - N_i - N_j) - E_{ij}(N_w)]  \nonumber \\
 & - &   [E_{ij}(N_w - N_i) - E_{ij}(N_w-N_j)] 
\end{eqnarray}
where $E_{ij}(N)$ is the probability of achieving $N$ electoral votes without either state $i$ or $j$.  The first line represents the combinatoric increase in eliminating states $i$ and $j$; the second line subtracts off the combinatoric increases with the exclusion of each state leaving only the effect of the correlation between states $i$ and $j$.

Within the thermodynamic analogy discussed above, $\chi_{ij}$ is the correlation function between state $i$ and state $j$.  The eigenvectors of $\chi_{ij}$ corresponding to large eigenvalues may be thought of as describing a set of states which contribute synergistically to either increasing or decreasing the probability of winning. Computational studies of this quantity are in progress at the time of this manuscript.

\begin{table}
\begin{tabular}{|lrr|}
\hline
        & Obama & \\
        State          & $\chi_j$ & $\chi_j / \sum$ \\ 
\hline 
        Ohio           & 0.161  & 0.141               \\ 
        Virginia       & 0.159  & 0.139               \\ 
        Nevada         & 0.155  & 0.136               \\ 
        Iowa           & 0.138  & 0.121               \\ 
        Wisconsin      & 0.117  & 0.103               \\ 
        Colorado       & 0.109  & 0.095               \\ 
        New Hampshire  & 0.095  & 0.083               \\ 
        Florida        & 0.087  & 0.076               \\ 
        North Carolina & 0.063  & 0.055               \\ 
        Michigan       & 0.058  & 0.051               \\
\hline
\end{tabular}
\caption{\label{Obama_table}Electoral susceptibilities (Obama), $\chi_j$ for ten swing states comprising $126$ electoral votes. The largest three susceptibility states (Ohio (18) , Virginia (13) and Nevada (6)) include a big, medium, and small electoral count state, showing the effects of interactions between states. }
\end{table}

\begin{table}
\begin{tabular}{|lrr|}
\hline
& Romney & \\
State          & $\chi_j$ & $\chi_j / \sum$ \\ 
\hline 
Ohio	&	0.159	&	0.150	\\
Virginia	&	0.154	&	0.145	\\
Wisconsin	&	0.116	&	0.109	\\
New Hampshire	&	0.116	&	0.109	\\
Nevada	&	0.115	&	0.108	\\
Colorado	&	0.104	&	0.098	\\
Iowa	&	0.104	&	0.097	\\
Florida	&	0.081	&	0.076	\\
Michigan	&	0.058	&	0.054	\\
North Carolina	&	0.057	&	0.054	\\
\hline
\end{tabular}
\caption{\label{Romney_table}Electoral susceptibilities (Romney), $\chi_j$ for ten swing states comprising $126$ electoral votes. The largest three susceptibility states (Ohio (18) , Virginia (13) and Wisconsin (10)) include a big, medium, and a relatively small electoral count state, showing the effects of interactions between states. }
\end{table}

\section{results for a $p_j=1/2$ electoral college.}

The susceptibility $\chi_j$ has been computed for a model with ten swing states (listed in Table (\ref{pequalshalf_table})) and $N_w =49$, corresponding to a common electoral scenario for the Democratic candidate entering the swing state electoral contest. The inner chart of Figure \ref{pequalshalf} show the unnormalized susceptibilities $\partial E/ \partial p_j$ for all of the states in the model. As expected, their shares are closely proportional to their electoral counts, $N_j$.  The normalized susceptibilities, $\chi_j$, shown in the outer chart, roughly equalizes the individual state shares.  

However close inspection of Table \ref{pequalshalf_table} shows that they are not exactly the same, the largest three being Florida, Nevada and Iowa.  These differences are a consequence of the subtle correlations between states present in the set of winning partitions.  For instance, even though Nevada (6) and New Hampshire (4) are of similar size, Nevada works cooperatively with more states and appears in more winning partitions (those with a vote total of $N_w =49$ or more).  Similarly, even though Ohio and Florida are the largest states, Ohio appears with a higher frequency in the winning partitions.  The presence of a correlation between two states that effects the states' susceptibility, $\chi_j$, depends very much upon $N_w$, a feature we attempt to explore analytically in section V.

\section{results for electoral college 2012}

Using aggregate polling data assembled at the FiveThirtyEight Blog from the beginning of October 2012 \cite{silver}, we have computed the electoral susceptibilities $\{ \chi_j\}$ of ten swing states. Specifically, the polling percentages for candidates and their variances are used in equation (\ref{normal}), which determine the probability, $p_j$, and parameter, $\alpha_j$, for state $j$. 

\begin{figure}[ht]
\includegraphics[width=7.5cm]{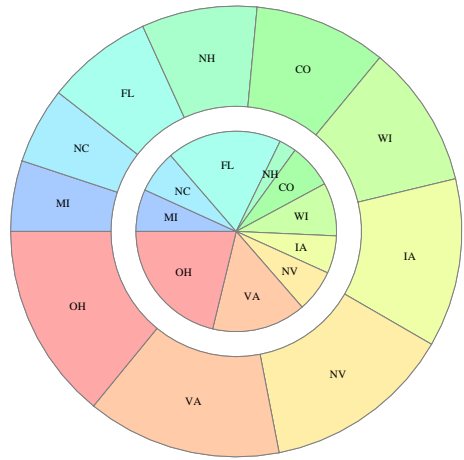}
\caption{\label{Obama} (Color online) Electoral susceptibility (Obama), $\chi_j$, for state $j$ displayed in relative terms (outer pie). Electoral susceptibility without normalization, $N_j \chi_j$, for state $j$ displayed in relative terms (inner pie.)}
\end{figure}

Table \ref{Obama_table}, first column, shows the electoral susceptibilities for President Obama listed in descending order.  Switching the voter equivalent of one electoral vote in the state of Ohio has the largest effect on his re-election probability compared with any other state.  It is significant to note that Virginia and Nevada have susceptibilities that are nearly as large even though (especially Nevada) they represent much smaller electoral shares.  The second column of Table \ref{Obama_table} is the same data normalized to sum to unity; this data is displayed in Figure (\ref{Obama}) to illustrate the relative impact of the ten swing states. Again, the inner pie of Figure (\ref{Obama}) shows the unnormalized susceptibilities which reflect the relative electoral vote contribution of state $j$, $N_j$, but not the effect of changing a fixed number of votes. 

\begin{figure}[ht]
\includegraphics[width=7.5cm]{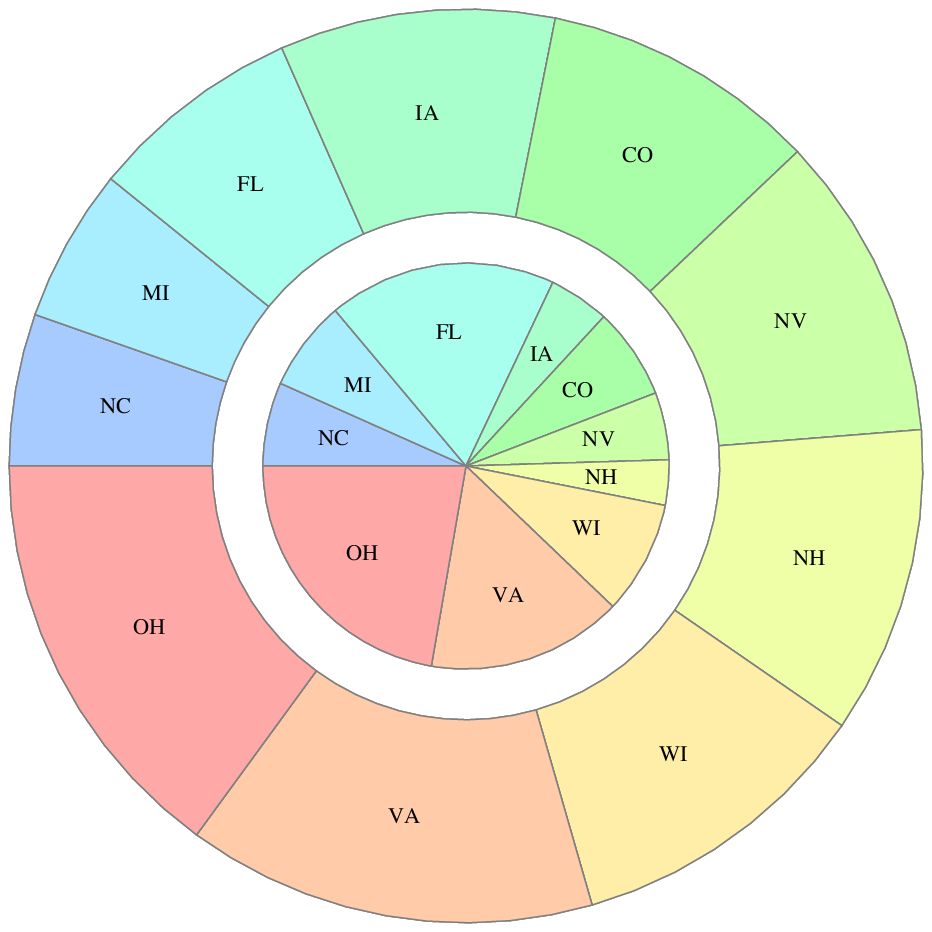}
\caption{\label{Romney} (Color online) Electoral susceptibility (Romney) $\chi_j$ for state $j$ displayed in relative terms (outer pie). Electoral susceptibility without normalization, $N_j \chi_j$, for state $j$ displayed in relative terms (inner pie.)}
\end{figure}

Table \ref{Romney_table} and Figure (\ref{Romney}) show the same analysis for Governor Romney.  Note that the susceptibilities for Obama and Romney are close but not exactly identical; this effect, mentioned in section II, is a consequence of an even number (126) of electoral swing state votes.

\begin{figure}[ht]
\includegraphics[width=7.5cm]{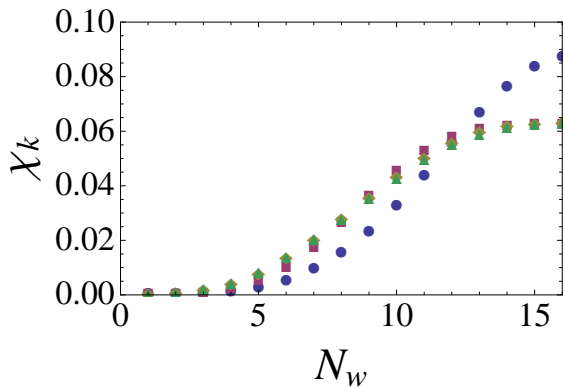}
\caption{\label{fig_Zipf} (Color online) Electoral susceptibility $\chi_k$ for a state at level $k=0 (\circ), 1 (\sqcap), 2 (\diamond), 3 ( \triangle)$.}
\end{figure}

\section{Zipf model}
\label{zipf}
Consider a model in which states with lower electoral counts are more numerous than states with higher electoral count.  We assume the states follow a Zipf distribution: there is a single state with electoral count $N_0$, two states with electoral count $N_0/2$; four states with electoral count $N_0/4$, etc. Denoting by $L_k$ the number of states at level $k$, and $N_k$ the electoral value of states at level $k$ the Zipf distribution is: $L_k = 2^k$ and $N_k = (1/2)^k N_0$. Let $q_k$ be the number of states won at level $k$.  The probability of achieving a particular set  $\{q_k\}$, analogous to equation (\ref{part_prob}) is 
\be
p(\{q_k\}) = \frac{1}{2^\Omega}\prod_k B^{L_k}_{q_k}
\ee
where $\Omega = \sum_k L_k$ is the total number of states and $B^n_m = \frac{n!}{m!(n-m)!}$ i is the binomial coefficient.  The probability of a candidate achieving the winning number of votes, $N_w$, or more (analogous to equation (\ref{prob})) is:
\be
E(N_w) = \sum_{\{q_k\}}{p(\{q_k\}) \theta(\sum_k{q_k N_k - N_w})}
\ee

For a model with four levels, the susceptibility of a state in level $k$, $\chi_k$, is shown in Figure \ref{fig_Zipf}.  Small $N_w$ corresponds  to a "one-sided" election where one candidate enters the swing state contest needing only a small number of votes to win, in this case $N_w << 32$, the total number of electoral votes.  For $N_w \sim 1$, it is possible to show that $\chi_k \propto 1/N_k$. As the election tends towards an "even" contest, $N_w \approx 16$, the susceptibilities becomes approximately constant $\chi_k \approx 1/N_w$ (independent of $k$ except for $k=0$)  and $\partial E /\partial p_k \approx N_k/N_w$.

\begin{figure}[ht]
\includegraphics[width=7.5cm]{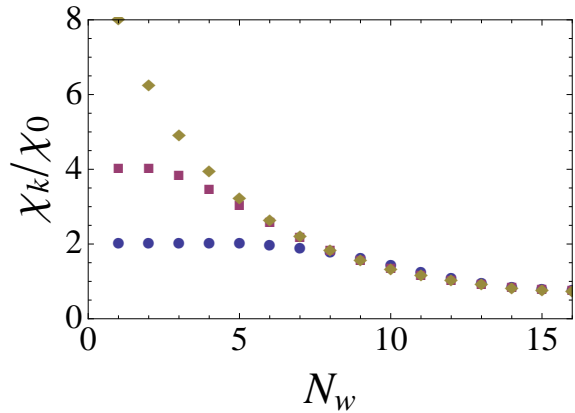}
\caption{\label{fig_Zipf_ratio} (Color online) Electoral susceptibility ratio $\chi_k/\chi_0$ for a state at level $k$. This is the susceptibility normalized with respect to the susceptibility of the single highest electoral count state. }
\end{figure}

For one-sided electoral contests, Figure \ref{fig_Zipf} shows that small electoral count states have the largest susceptibilities; as the contest becomes more even, the largest electoral count state becomes dominant with all other states tending toward equal susceptibilities. Figure \ref{fig_Zipf_ratio}, which normalizes the susceptibility of a state at level $k$ with that of level $k=0$ illustrates this trend. 

\section{Conclusion}

The Electoral Susceptibility $\chi_j$ is roughly a measure of where campaign capital to change votes should be spent to maximize the probability of winning the election.  Our analysis points to subtle correlations between the frequency of states' appearances in the set of winning partitions of the electoral college which is not easily intuited.  

From the analysis of the hierarchical Zipf model (section \ref{zipf}) it is seen that in close swing state electoral contests, the largest state has the largest susceptibility, and in "one-sided" contests small states have the highest susceptibilities.  

The latter effect may be partially understood simply by looking at the dependence of the winning probability, $E(N_w)$, upon the average state size in electoral votes.  For instance in the example given in the introduction of ten swing states, the ratio of the number of Democrat winning partitions to Republican winning partitions was $741/283 \approx 2.62$.  The ratio of winning electoral votes (Republican/Democrat) was $79/49 \approx 1.61$. If the election were an "even" electoral contest, both ratios would be unity.  The proliferation of winning to losing partitions compared with the ratio of electoral votes is a reflection of the number of relatively small states giving rise to large number of winning combinations for the candidate (the Democrat in this case) who enters the swing state electoral contest in the lead. Consider the extreme example: a candidate needing only 5 electoral votes would much prefer an electoral college having 20 states carrying 1 vote apiece than an electoral college having 5 states carrying 4 votes apiece.  Small states are more valuable per capita because they carry a larger per capita entropy.

However the situation is more complex in the realistic, highly inhomogeneous, electoral college. Even in the $p_j =1/2$ equal probability case, several large and small states have comparable susceptibilities. The 49 votes needed by the Democratic candidate corresponds to neither a close nor "one-sided" electoral contest. Also, from the analysis above it is not obvious why the susceptibilities are exactly the same for the leading {\sl and} trailing candidate.

Lastly, we mention how this calculation might be improved for a realistic campaign.  It is well known that one electoral vote is equivalent to a different number of voters in different states.  (For up to date demographics see reference \cite{demographics}.)  This correspondence needs to be folded through our derivative leading to equation \ref{dpds}.

This research was supported by the Department of Energy DE-FG02-08ER64623---Hofstra University Center for Condensed Matter, Research Corporation CC6535 (GL), the Howard Hughes Medical Institute Scholar Program (BC) and the Columbia University program on the Genetics of Complex Disorders (JC).


\begin{references}
\bibitem{ev} The parameter, $s_j$, corresponds to population equivalent to one electoral vote in state $j$.  Of course, the demographics of the electoral college system make the population equivalent to one electoral vote different in every state.  Our calculation could be easily modified to include this effect.
\end{references}
\end{document}